\documentclass[cits]{PoS}

\title{Polarization in Pulsar Wind Nebulae}

\ShortTitle{Polarization in PWNe}

\author{\speaker{D. Volpi} $^a$, L. Del Zanna $^b$, E. Amato $^c$, and N. Bucciantini $^d$  \\
\llap{$^a$ $^b$}Dipartimento di Astronomia e Scienza dello Spazio-Universit\'a degli Studi di Firenze, Largo E. Fermi, 2, 50125, Firenze, Italy\\
\llap{$^c$}INAF-Osservatorio Astronomico di Arcetri, Largo E. Fermi 5, 50125, Firenze, Italy \\
\llap{$^d$}Astronomy Department, University of California at Berkeley,
601 Campbell Hall, Berkeley, CA 94720-3411, USA\\
\email{delia@arcetri.astro.it}, \email{ldz@arcetri.astro.it}, \email{amato@arcetri.astro.it}, \email{nbucciantini@astro.berkeley.edu}}

\abstract{The main goal of our present work is to provide, for the first time, a simple computational tool that can be used to compute the brightness, the spectral index, the polarization, the time variability and the spectrum of the non-thermal light (both synchrotron and inverse Compton, IC) associated with the plasma dynamics resulting from given relativistic magnetohydrodynamics (RMHD) simulations. The proposed method is quite general, and can be applied to any scheme for RMHD and to all non-thermal emitting sources, e.g. pulsar wind nebulae (PWNe), and in particular to the Crab Nebula (CN) as in the present proceeding. Here only the linear optical and X-ray polarization that characterizes the PWNe synchrotron emission is analyzed in order to infer information on the inner bulk flow structure, to provide a direct investigation of the magnetic field configuration, in particular the presence and the strength of a poloidal component, and to understand the origin of some emitting features, such as the knot, whose origins are still uncertain. Future work will be necessary to obtain polarization maps in the hard X- and gamma-rays to compare with the current observations by Integral/IBIS \cite{forot}. The complete investigation of the dynamics and synchrotron emission can be found in \cite{delzannac}. In \cite{volpi} the inverse Compton radiation is examined to disentangle the different contributions to radiation from the magnetic field and the particle energy distribution function, and to search for a possible hadronic component in the emitting PWN, and thus for the presence of ions in the wind. If hadronic radiation was found in a PWN, young supernova remnants would provide a natural birth-place of the cosmic-rays (CRs) up to the so-called \emph{knee} in the CR spectrum.}

\FullConference{Polarimetry days in Rome: Crab status, theory and prospects\\
		 October 16-17 2008\\
		 Rome, Italy}

\begin{document}

\section{Synchrotron emission and polarization}
PWNe are a class of supernova remnants that is characterized by a non-thermal continuum emission from the radio up to soft gamma-ray band mainly due to synchrotron emission, and in the gamma-rays principally due to IC scattering. The optical and X-ray observations made by the Hubble Space Telescope, ROSAT, Chandra, and XMM-Newton show that the nebular continuum emission presents a jet-torus morphology. Some examples are the CN, which is the prototype of PWNe  \cite{hestera}, \cite{weisskopfa}, \cite{hesterb}, and the nebula associated with Vela \cite{pavlova}, \cite{pavlovb}.

Analytical uni-dimensional models were created to explain the physics at the base of the PWNe \cite{kennela}, \cite{kennelb}, \cite{emmering}, but they are inadequate to describe the jet-torus structure. Analytical bi-dimensional models thus were born in order to give an answer to the problem \cite{bogovalov}, \cite{lyubarsky}. The idea was that an anisotropic Poynting plus kinetic energy flux in the pulsar wind (PW) was capable of giving origin to the torus and the oblate shape of the termination shock (TS, the shock in which the PW ends). The jets appear to come out from the pulsar due to the cusp-like form of the shock, instead they are collimated by the hoop-stresses of the nebular magnetic field downstream of the TS. This idea was confirmed by RMHD simulations \cite{komissarova}, \cite{delzannaa}, which were able to solve the bi-dimensional hyperbolic equations thanks to the developed numerical codes \cite{komissarovb}, \cite{delzannab}. 

In order to investigate the PWN dynamics and emission it is crucial to study the polarization. As a matter of fact the synchrotron emission from relativistic particles is known to be linearly polarized ($V_{\nu}=0$) with a high degree of polarization. The polarization probes geometry of the source and the magnetic field strength and direction, together with the particle acceleration.

In our numerical model the dynamics and the synchrotron emission of the PWNe are deduced from simulations obtained from a two-dimensional axisymmetric RMHD shock-capturing code \cite{delzannaa}, \cite{delzannab}. In the code the maximum particle energy ($\epsilon_\infty$) equation together with the RMHD equations, is evolved in space and time following \cite{delzannac}
\begin{equation}
\frac{d\epsilon_\infty}{dt^{\prime}}=\frac{d \ln{n^{1/3}}}{dt^{\prime}}+\frac{1}{\epsilon_\infty}(\frac{d\epsilon_\infty}{dt^{\prime}})_{\mathrm{sync}}
\end{equation}
where the first and the second terms of the second member are respectively the adiabatic and the synchrotron losses, $t^{\prime}$ is the time in the comoving frame, and $n$ is the proper numerical density. 
Our model has three parameters: the parameter $\alpha$ connected to the wind energy flux anisotropy, the parameter $b$ linked to the width of the striped wind region, and the magnetization $\sigma$.

For the set of parameters called runA (a factor of $10\%$ in anisotropy, $\sigma_{\mathrm{eff}}=0.02$, and a narrow striped wind region) the CN dynamics is well reproduced: both the bulk flow speed, and the jet-torus morphology are very similar to the observations. 

A particle-energy distribution function considered as a power law both at the injection and evolved along the streamlines together with a monochromatic-approximated spectral power, allows the emission coefficient, in which the relativistic corrections are included, to be calculated
\begin{equation} 
j_{\nu}(\nu,\vec{n}) = \left\{\begin{array}{ll}
C P \,|\vec{B}^\prime\times\vec{n}^\prime|^{a +1}
D^{a +2}\nu^{-a}, & \nu_\infty\ge\nu, \\
 & \\
0, & \nu_\infty < \nu. \end{array}\right.
\end{equation}
where $C$ is a constant; $a$ is the spectral index; $P$ is the thermal pressure; $\vec{B}^\prime=B/\gamma$ is the magnetic field in the comoving frame with $B$ the magnetic field in the observer frame and $\gamma$ the fluid Lorentz factor; $\vec{n}^\prime$ and $\vec{n}$ are the observer direction respectively in the comoving and in the observer frame; $\nu$ is the observation frequency; and $D$ is the Doppler boosting factor.
The synchrotron burn-off is reproduced by a cut-off frequency that is the maximum particle frequency $\nu_{\infty} \equiv D \nu_c^\prime (\epsilon_\infty) =
D\,3e/(4\pi mc)\,|\vec{B}^\prime\times\vec{n}^\prime|
\epsilon_\infty^2$.

From the emissivity it is possible to calculate the surface brightness and the spectral index maps, the integrated spectra, the Stokes parameters, and the degree and direction of linear polarization \cite{delzannac}.

\begin{figure*}[!t]
\centerline{\resizebox{0.6\hsize}{!}{
\includegraphics{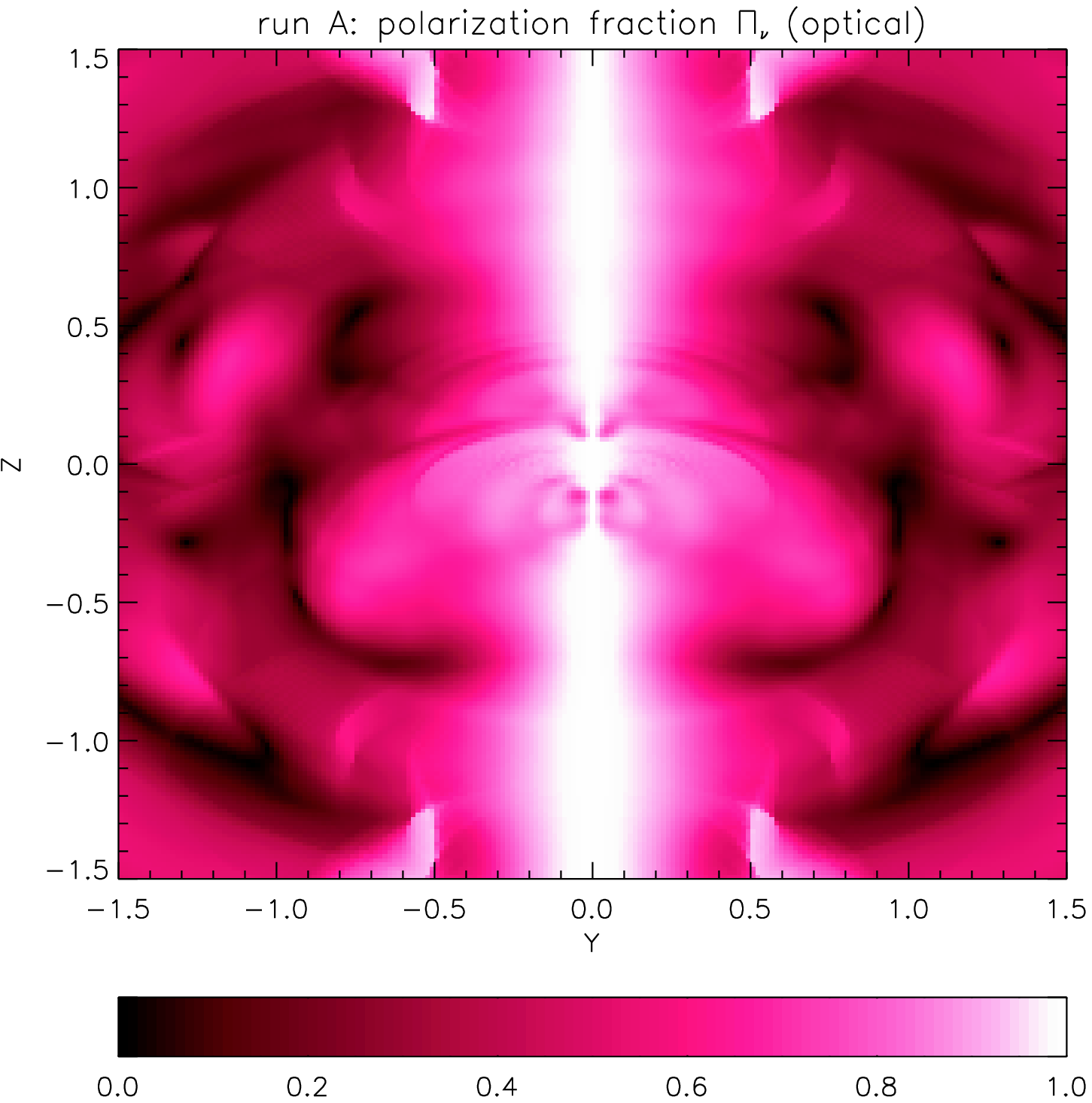}
\includegraphics{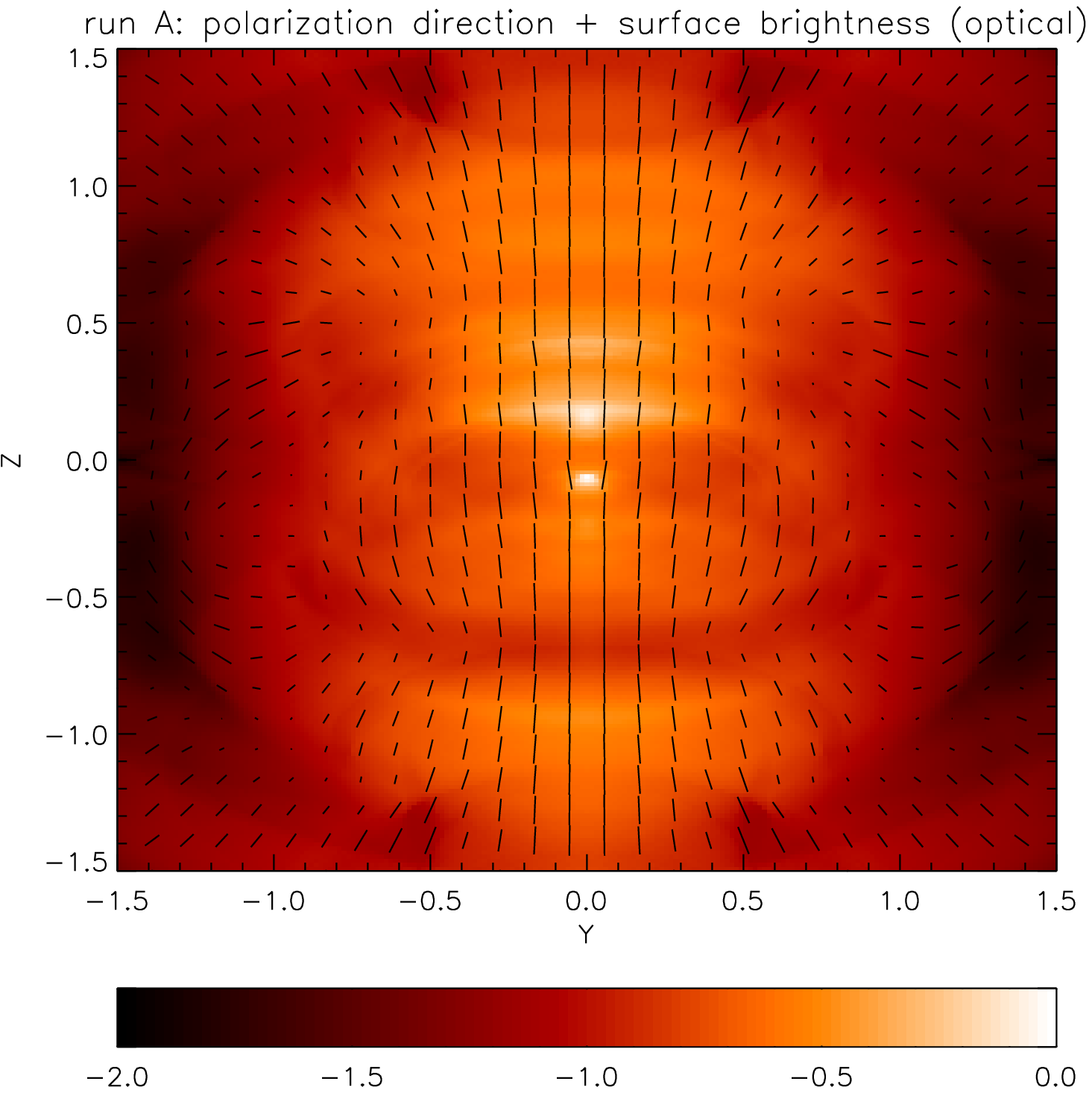}
}}
\centerline{\resizebox{0.6\hsize}{!}{
\includegraphics{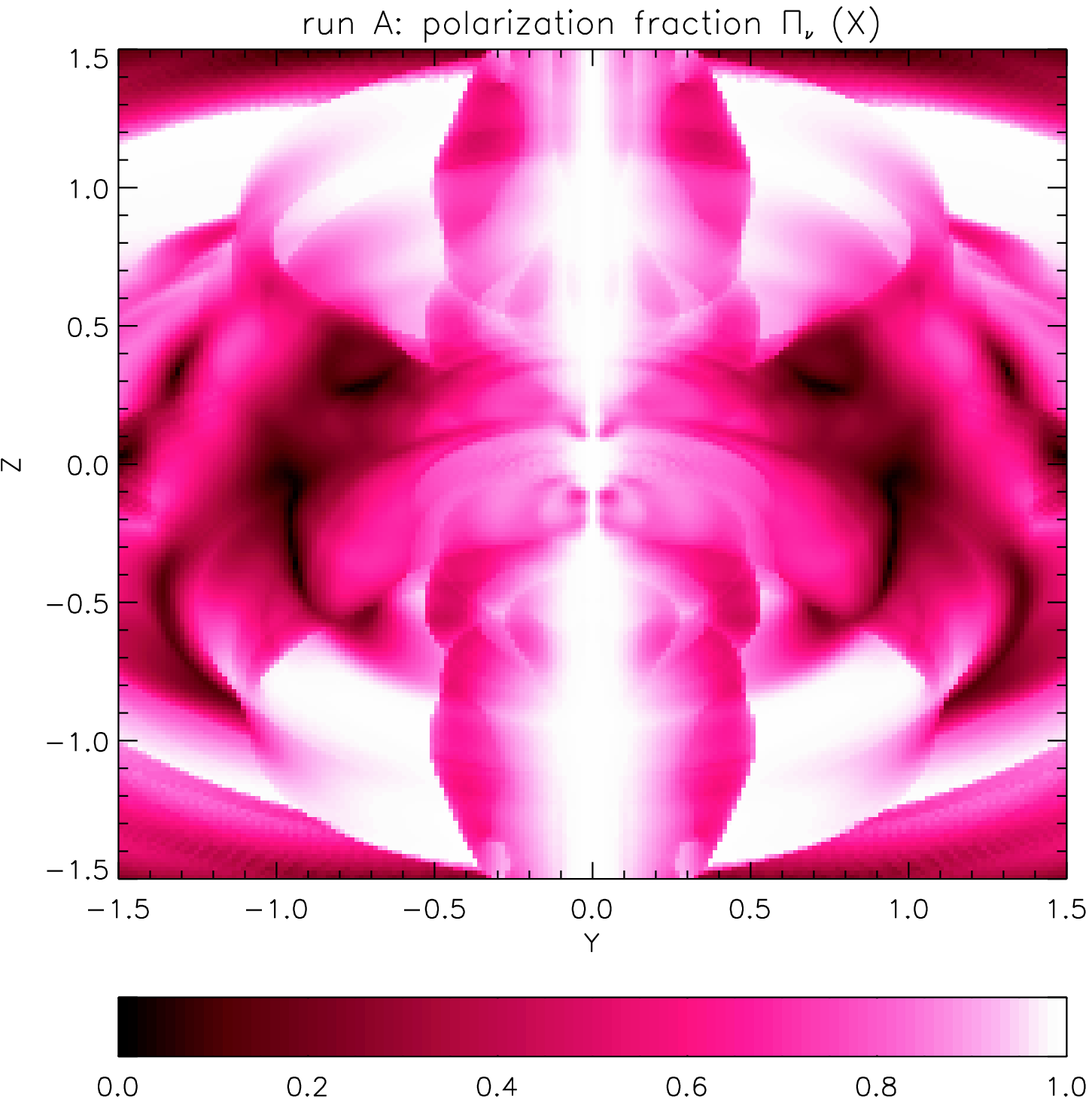}
\includegraphics{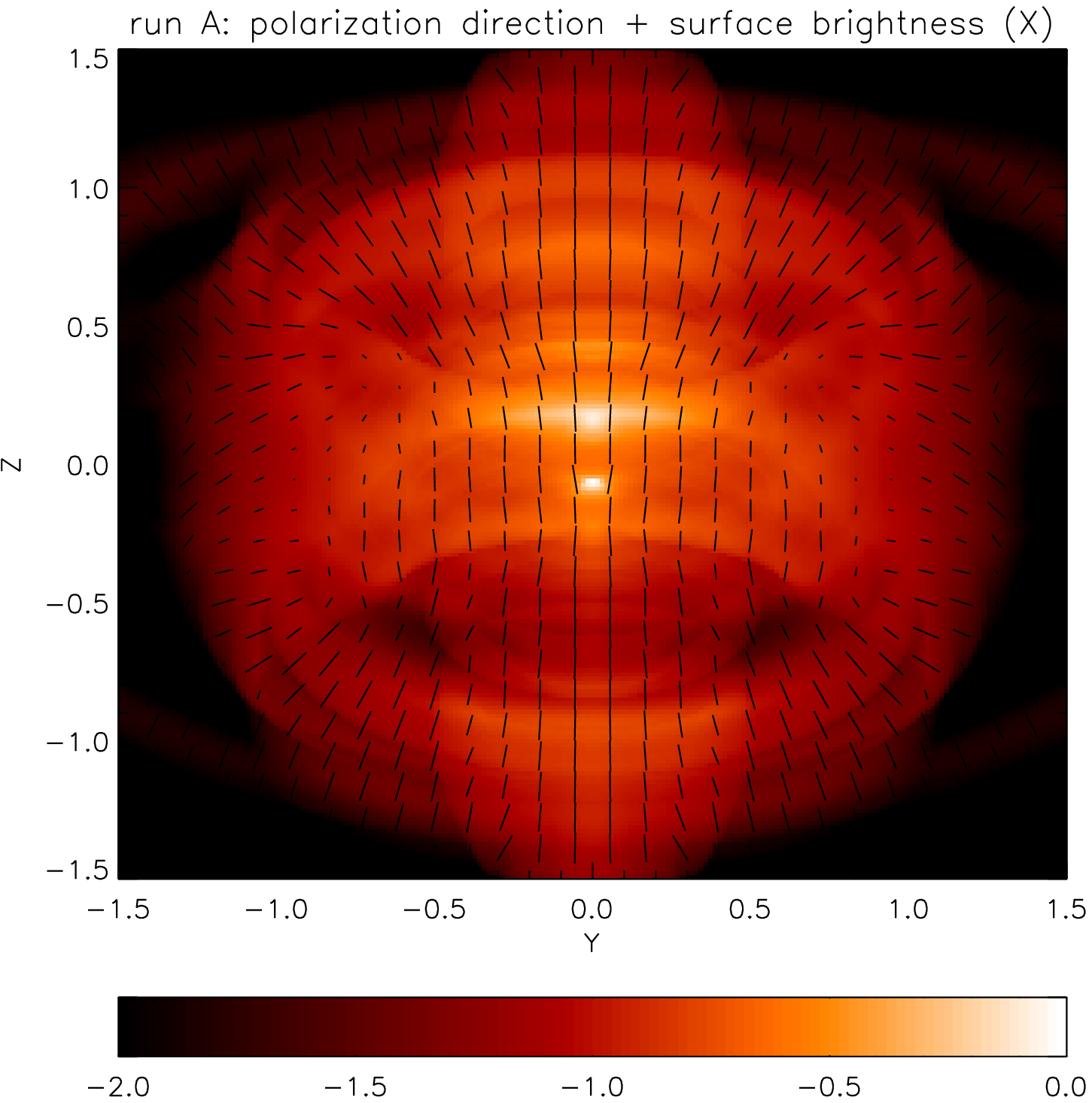}
}}
\caption{
\small{Maps of the polarization fraction $\Pi_\nu$ (left) and direction 
(right, superimposed on the surface brightness map), 
at the optical frequencies (upper row) and in the X-rays (lower row) for runA. Polarization fraction is normalized against $(a+1)/(a+5/3)\simeq 70\%$ for a spectral index $a=0.6$ selected from optical observations of the Crab Nebula inner region \cite{veron}. The tick 
length that is used for polarization direction is proportional to the normalized 
fraction. The emission (logarithmic scale) is normalized to the maximum, see the color panel. The inclination angle of the symmetry axis with respect to the plane of the sky is $\alpha=30^\circ$ as in the Crab Nebula, while the tilt is retained. Along the axes the distance from the central pulsar is in ly.}
}
\label{fig:polarization}
\end{figure*}

In particular the surface brightness in the cartesian coordinate system ($x, y, z$) is given by
\begin{equation} 
I_\nu (\nu,y,z) = \int_{-\infty}^{\infty}j_\nu (\nu,x,y,z)\,\mathrm{d}x.
\end{equation}
The corresponding synthetic optical and X-ray maps appear to reproduce the jet-torus structure, and the size reduction due to the synchrotron burn-off at increasing frequencies. Our maps are qualitatively comparable to the observed images even in the finest details such as the inner ring, the central knot and the main arc (see the right panel of Fig.\,\ref{fig:polarization} for the runA case that best matches the CN dynamics). Also the synthetic spectral index maps are very similar to those of Mori \cite{mori}.

The Stokes parameters $Q_\nu$ and $U_\nu$ are instead obtained from the following equations
\begin{equation} 
Q_\nu (\nu,y,z) = \frac{a+1}{a+5/3}\int_{-\infty}^{\infty}
j_\nu (\nu,x,y,z)\cos 2\chi\,\mathrm{d}x,~~~~~~U_\nu (\nu,y,z) = \frac{a+1}{a+5/3}\int_{-\infty}^{\infty}
j_\nu (\nu,x,y,z)\sin 2\chi\,\mathrm{d}x,
\end{equation}
where $\chi$ is the local polarization position angle between the emitted electric field vector $\vec{e}$ in the plane of the sky, measured from the z-axis; and the additional factor $ (a+1)(a+5/3)$ comes from the intrinsic properties of synchrotron emission.

Finally the polarization fraction ($\Pi$), and direction ($P$) can be thus deduced as
\begin{equation} 
\Pi_\nu=\frac{\sqrt{Q_\nu^2+U_\nu^2}}{I_\nu}~~~~~~~~~~~~~~~~ \vec{P_\nu}=\Pi_\nu \left( \sin{\chi}\hat{x}+\cos{\chi}\hat{y}\right).
\end{equation}

\begin{figure*}[!t]
\centerline{\resizebox{0.8\hsize}{!}{
\includegraphics{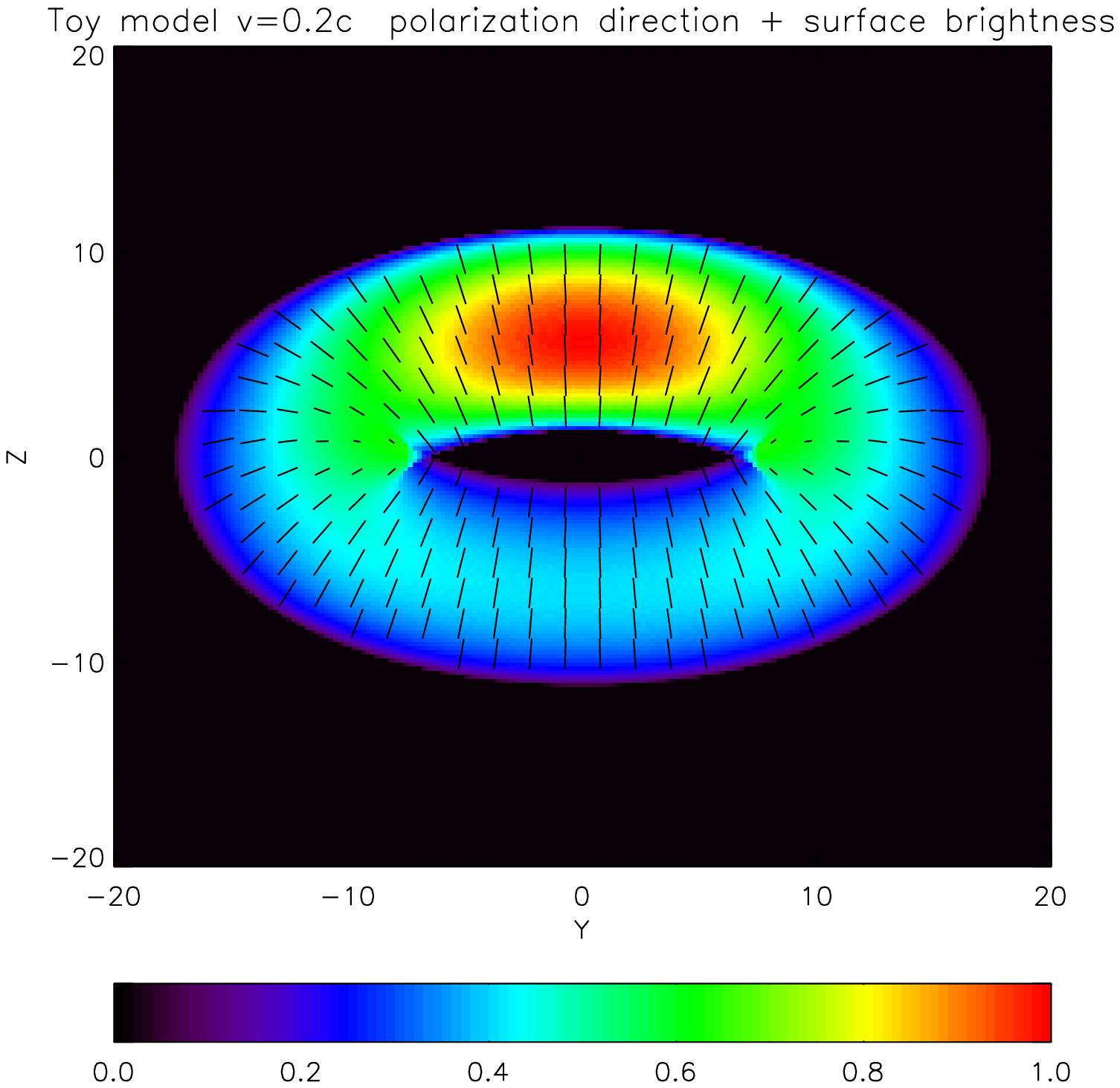}
\includegraphics{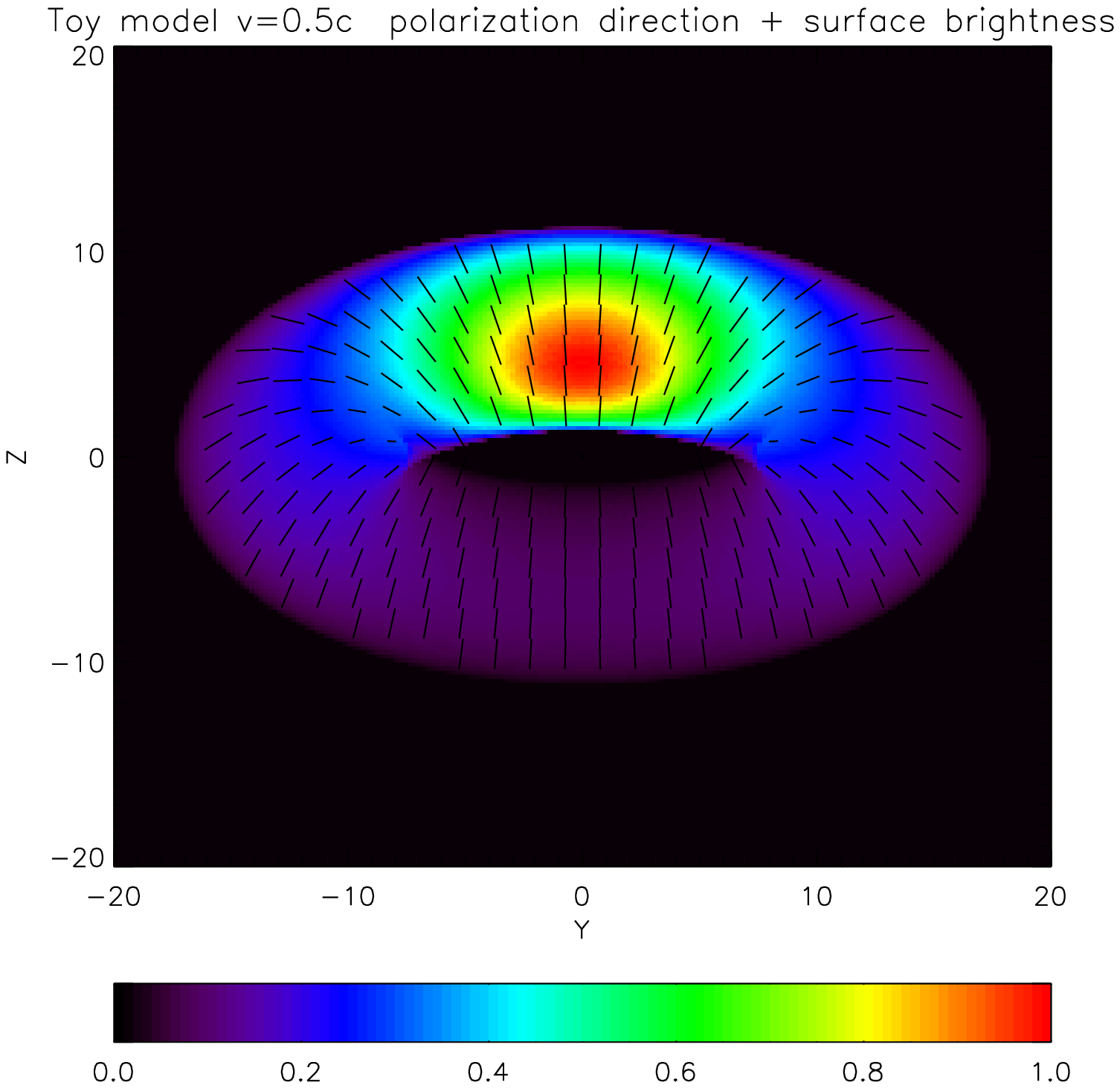}
\includegraphics{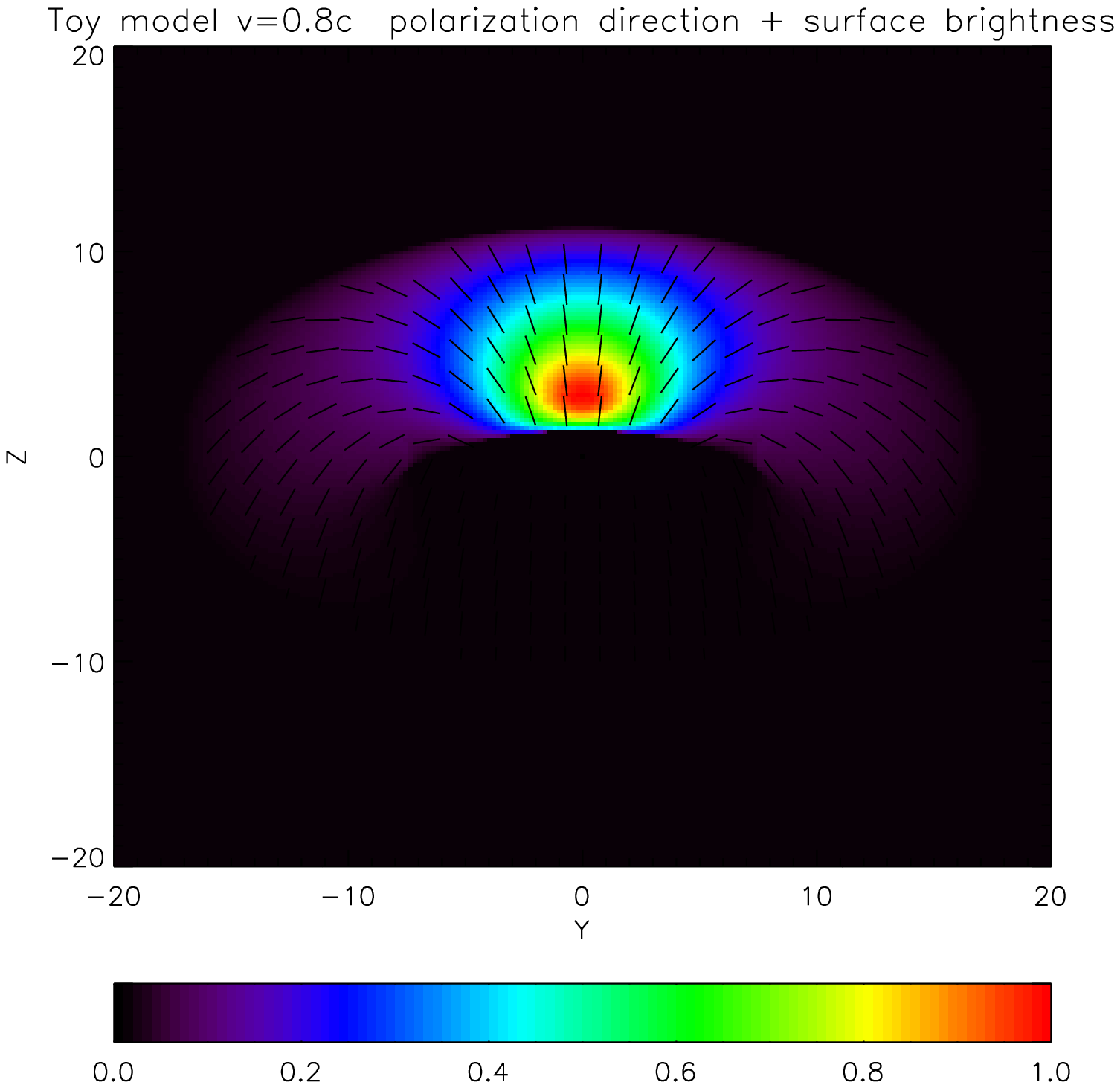}
}}
\caption{
\small{Maps of polarization direction (superimposed to the surface brightness map) in the uniform torus case at increasing values of the velocity: on the left 0.2~c; in the middle 0.5~c; on the right 0.8~c. Polarization fraction is normalized against $(a+1)/(a+5/3)\simeq 70\%$ for a spectral index $a=0.6$ selected from optical observations of the Crab Nebula inner region \cite{veron}. The tick length that is used for polarization direction is proportional to the normalized fraction. The emission (linear scale) is normalized to the maximum, see the color panel. The inclination angle of the symmetry axis with respect to the plane of the sky is $\alpha=30^\circ$ as in the Crab Nebula, while the rotation angle towards West with respect to the North is zero. Along the axes the distance from the center of the torus is in arbitrary units.}
}
\label{fig:polarization_toy}
\end{figure*}
The polarization formulae and their results are tested and investigated first through an uniform emitting torus as in \cite{bucciantini}, and then through the Kennel and Coroniti model, see \cite{nakamura}, in order to understand the polarization behavior. The $\Pi_{\nu}$ results are reported in figure \ref{fig:polarization_toy} for three bulk flow speeds. The flow speeds are taken to be greater than 0.2 c in order to include significant Doppler boosting effects. The polarization vector that is close to the symmetry axis of the torus is aligned with the axis itself due to the assumption of a purely toroidal magnetic field. Moving away from the symmetry axis, the inclination of the polarization vector with respect to the same axis increases.  The position where the polarization vector is perpendicular to the symmetry axis is a function of the flow velocity for a constant torus inclination angle. It is immediately evident that for increasing values of the bulk flow speed the corresponding position at which $\Pi_{\nu}$ deviates from parallelism to the symmetry axis, is closer and closer to the aforesaid axis. The position of the transition can be measured with the polar angle $\theta$, the results are: $90^{\circ}$, $75^{\circ}$, and $60^{\circ}$ respectively for a bulk flow speed of 0.2 c, 0.5 c, and 0.8 c. This effect is called the polarization angle swing (thus the deviation of the vector direction) and is due only to relativistic effects. The angle swing is bigger in the front side than in the back side of the torus, as observed \cite{schmidt}, with an increasing discrepancy for increasing flow speed values. The $\Pi_{\nu}$ images are thus capable of giving information about the flow speed. The polarization fraction maps are not interesting, because they reach the maximum theoretical value almost overall the torus, regardless of the radial speed.

Figure \ref{fig:polarization} shows $\Pi_{\nu}$ on the left and $P$ on the right in the optical ($5.6 \times 10^{4}$\,Hz, upper panels), and in the X-ray (1 keV, lower panels) frequencies for the runA inner region. The $\Pi_{\nu}$ images show a higher polarized fraction along the polar axis, where the projected magnetic field $B$ (assumed to be toroidal) is always orthogonal to the line of sight, and in the central region of the torus especially around the front side of the bright arcs, while depolarization occurs in the outer regions, where contributions from the projected fields with opposite signs sum up along the line of sight. The polarization direction $P$ shows several ticks with a length proportional to $\Pi_\nu$, superimposed on the surface brightness maps, with the aim of making clear the association between the polarization behavior and the emission main features. The polarization ticks are basically always orthogonal to $B$, displaying the behavior expected given the inclination of the symmetry axis with respect to the plane of the sky. However, in the rings where the velocity is relativistic, deviations of the vector direction due to polarization angle swing are also visible. From the inner ring up to the external arcs in the torus the polarization vector becomes perpendicular to the symmetry axis of the nebula for growing values of the polar angle $\theta$ due to the decreasing bulk flow speed. In the back side of the inner ring (or of the arcs) the polarization tends to remain parallel to the axis up to larger angular distances than in the front side. This asymmetry could be used to infer the relativistic speeds in the inner parts of the nebula. If an isotropic magnetic field is considered, the growth of the depolarized areas in the polarization fraction maps are slightly visible only at smaller scales, while the variations in the polarization direction maps result negligible. Notice that in these zoomed maps it is possible to see a small bright feature resembling the CN knot (displaced to the South with respect to the central position,  as in the figures by \cite{hestera}) that in our simulations seems to originate in the cusp-like region immediately downstream of the oblate TS. The behavior of the polarization direction in the internal region of the torus can give information about the flow speed, while the polarization fraction could estimate the amount of disordered or poloidal magnetic field. A comparison with the observations at the corresponding frequencies is not possible, because high resolution optical and X-ray polarization maps of the inner region of the Crab Nebula, or of other PWNe, which could really provide crucial clues to the magnetic structure, are not available yet. X-ray observations were made by Weisskopf at $2.6$~keV and $5.2$~keV \cite{weisskopfb}, but they were not spatially resolved. Weisskopf found a mean polarization fraction of $\approx 20\%$ at both frequencies, which is the same value obtained in the optical band \cite{oort}, and a polarization angle of $\approx 156^{\circ}$. The run A averaged polarization fractions agree with the corresponding observed values (we find $\approx 18\%$ at both optical and X-ray frequencies), while the synthetic mean polarization angle would be reliable only in three-dimensional simulations.

\end{document}